\documentclass[a4paper,12pt]{article}
\usepackage{paper}

\title{A Real-Time Spatial Index for In-Vehicle Units}
\author{Magnus Lie Hetland}
\affil{Norwegian University of Science and Technology\\\url{mlh@idi.ntnu.no}}
\author{Ola Martin Lykkja}
\affil{Q-Free ASA, Trondheim, Norway\\\url{ola.lykkja@q-free.com}}

\begin{document}

\maketitle

\begin{abstract}
We construct a spatial indexing solution for the highly constrained
environment of an in-vehicle unit in a distributed vehicle tolling scheme
based on satellite navigation (\textsc{gnss}). We show that a purely
functional implementation of a high-fanout quadtree is a simple, practical
solution that satisfies all the requirements of such a system.
\end{abstract}

\section{Introduction}

Open road tolling is an increasingly common phenomenon, with transponder-based
tolling proposed as early as in 1959~\citep{Kelly:2006}, and a wide variety of
more complex technologies emerging over the recent decades~\citep{Hills:2002}.
One of the more recent developments is the use of satellite navigation
(\textsc{gnss}), with geographical points and zones determining the
pricing~\citep[see, e.g.,][]{Grush:2008}. In this paper, we examine the
feasibility of maintaining a geographical database in an in-vehicle unit,
which can perform many of the tasks of such a location-based system
independently.

This is a proof-of-concept application paper. Our main contributions can be
summed up as follows:
\begin{enumerate*}[label=(\textit{\roman*}), itemjoin={{; }},
        itemjoin*={{; and }}]
    \item We specify a set of requirements for a spatial database to be used in
        the highly constrained environment of a real-time, low-cost in-vehicle
        unit in \cref{sec:problem}
    \item we construct a simple data structure that satisfies these
        requirements in \cref{sec:solution}
    \item we tentatively establish the feasibility of the solution through an
        experimental evaluation in \cref{sec:exp}.
\end{enumerate*}
In the interest of brevity, some technical details have been omitted. See the
report by \citet{Lykkja:2012} for more information.

\section{The Problem: Highly Constrained Spatial Indexing}\label{sec:problem}

The basic functionality of our system is to retrieve relevant geographic
(i.e., geometric) objects, that is, the tolling zones and virtual toll
gantries that are within a certain distance of the vehicle. Given the
real-time requirements and the limited speed of the available hardware, a
plain linear scan of the data would be infeasible even with about 50 objects.
A real map of zones and gantries would hold orders of magnitude more objects
than that (c.f., \cref{tab:perf}). This calls for some kind of geometric or
spatial indexing~\citep{Samet:2006}, although the context places some heavy
constraints on the data structure used. One fundamental consideration is the
complexity of the solution. In order to reduce the probability of errors, a
simple data structure would be preferable. Beyond simplicity, and the need for
high responsiveness, we have a rather non-standard hardware architecture to
contend with.

\begin{figure}
\def\Scale{1.0}
\centering
\scalebox{\Scale}{\begin{tikzpicture}
    \fontsize{10}{10pt}\selectfont

    \foreach \x in {0,5} {
                \draw[shift={(\x,0)}, densely dotted]
            (3,-1) -- (5,0) -- (5,-4) -- (3,-2) -- cycle
            ;
    }
    \foreach \n/\c/\x in {Sector/k/0,Subsector/16/5,Page/16/10} {

        \draw[shift={(\x,-4)}]

            (0,0) rectangle (3,4)

            \foreach \y in {1,2,3} {
                (0,\y) -- (3,\y)
            }

            (1.5,0.5) node {\n\ $\c$ of $\c$}
            (1.5,1.5) node {$\cdots$}
            (1.5,2.5) node {\n\ 2 of $\c$}
            (1.5,3.5) node {\n\ 1 of $\c$}

            ;

        \draw[shift={(\x,-4)}, decorate, decoration=brace]
            (3.2,3) -- (3.2,2)
            ;

    }

    \draw (3.3,-1.5) node[right] {\strut\SI{64}{\kibi\byte}};
    \draw (8.3,-1.5) node[right] {\strut\SI{4}{\kibi\byte}};
    \draw (13.3,-1.5) node[right] {\strut\SI{256}{\byte}};
    \draw (1.5,-4.5) node {\strut$k=16\ldots 64$};

\end{tikzpicture}}
\caption{Flash memory architecture}\label{fig:architecture}
\end{figure}
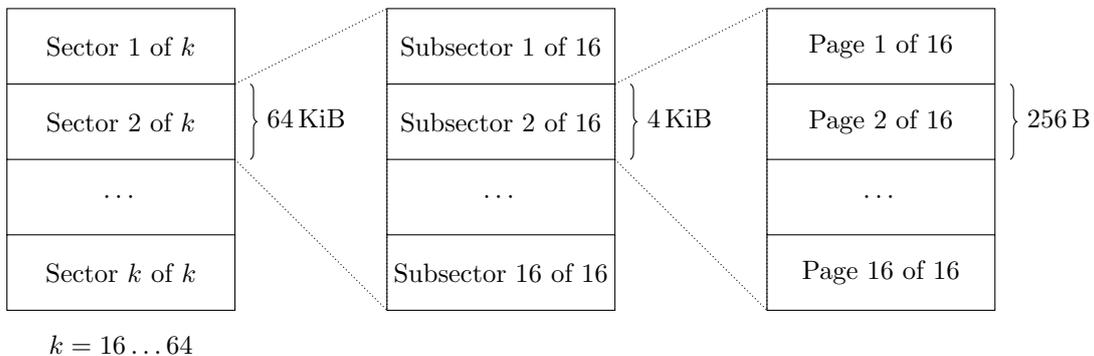

The memory\label{p:memarch} of the on-board unit is assumed to be primarily
flash memory with serial access. The scenario is similar to that of a desktop
computer, where the index would be stored on a hard drive, with a subset in
RAM and the L2 cache. For an overview of the hierarchical nature of the
memory architecture, see \cref{fig:architecture,tab:architecture}.
The serial nature of the memory forces us to read and write single bits at a
time in a page. A read operation takes \SI{50}{\micro\second}. A write
operation takes \SI{1}{\milli\second}, and may only alter a bit value of~1 to
a bit value of~0. A sector, subsector or page may be erased in a single
operation, filling it with 1-bits in approximately $\SI{500}{\milli\second}$.
One important constraint is also that each page may typically only be erased a
limited number of times (about \num{100000}), so it is crucial that our
solution use the pages cyclically, rather than simply modifying the current
ones in-place, to ensure wear leveling.

\begin{table}
\caption{Hierarchical memory architecture}\label{tab:architecture}
\begin{tabularx}{\textwidth}{@{}Xlll@{}}
\toprule
\bf Description & \bf Size & \bf Access time & \bf Persistent? \\
\midrule
Processor internal memory & \si{\kibi}-bytes & Zero wait state & No \\
External RAM & \SI{100}{\kibi\byte} & Non-zero wait states & No \\
Serial Flash & 8, 16 or \SI{32}{\mebi\byte} & See main text & Yes \\
\bottomrule
\end{tabularx}
\end{table}

Based on general needs of a self-contained tolling system, and on the hardware
capabilities just described, we derive the following set of requirements:

\begin{reqs}

\item\label{req:vers} The system must accommodate multiple versions of the
    same index structure at one time. The information may officially change at
    a given date, but the relevant data must be distributed to the units ahead
    of time.
\item\label{req:incr} It must be possible to distribute new versions as
    incremental updates, rather than a full replacement. This is crucial in
    order to reduce communication costs and data transfer times. It will also
    make it less problematic to apply minor corrections to the database.

\item\label{req:foot} The memory footprint must be low, as the memory
    available is highly limited.

\item\label{req:itgr} The database must maintain \SI{100}{\percent} integrity,
    even during updates, to ensure uninterrupted access. It must be possible
    to roll back failed updates without affecting the current database.

\item\label{req:cpu} The indexing structure must be efficient in terms of CPU
    cycles during typical operations. Both available processing time and
    energy is highly limited and must not be wasted on, say, a linear scan
    over the data.

\item\label{req:flash} The relevant data structure must minimize the number of
    flash page reads needed to perform typical operations, especially for
    search. Flash-page reads will be a limiting factor for some operations, so
    this is necessary to meet the real-time operation requirements.

\item\label{req:wear} In order to avoid overloading individual pages (see
    discussion of memory architecture, above), wear leveling must be ensured
    through cyclic use.

\item\label{req:oper} The typical operations that must be available under
    these constraints are basic two-dimensional geometric queries (finding
    nearby points or zones) as well as modification by adding and removing
    objects.

\end{reqs}
In addition, the system must accommodate multiple independent databases, such
as tolling information for adjacent countries. Such databases can be
downloaded on demand (e.g., when within a given distance of a border), cached,
and deleted on a least-recently-used basis, for example. We do not address
this directly as a requirement, as it is covered by the need for a low memory
footprint per database (\cref{req:foot}).

As in most substantial lists of requirements, there are obvious synergies
between some (e.g., \cref{req:cpu,req:flash}) while some are orthogonal and
some seem to be in direct opposition to each other (e.g.,
\cref{req:vers,req:foot}). In \cref{sec:solution}, we describe a simple index
structure that allows us to satisfy all our requirements to a satisfactory
degree.

\section{\boldmath Our Solution: Immutable 9-by-9
Quadtrees}\label{sec:solution}

The solution to the indexing problem lies in combining two well-known
technologies: quadtrees and immutable data structures.

In the field of geographic and geometric databases, one of the simplest and
most well-known data structures is the quadtree, which is a two-dimensional
extension of the basic binary search tree. Just as a binary search tree
partitions $\mathbb{R}$ into two halves, recursively, the quadtree partitions
$\mathbb{R}^2$ into quadrants. A difference between the two is that where the
binary search tree splits based on keys in the data set, the quadtree computes
its quadrants from the geometry of the space itself.\footnote{Technically,
this is the form of quadtrees known as PR
Quadtrees~\citep[\S\,1.4.2.2]{Samet:2006}.} In order to reduce the number
of node (i.e., page) accesses, at the expense of more coordinate
computations, we increase the grid of our tree from 2-by-2 to
9-by-9. The specific choice of this grid size is motivated by the
constraints of the system. We need 3 bytes to address a flash page and
with a 9-by-9 grid, we can fit one node into $9\cdot 9\cdot 3=243$
bytes, which permits us to fit one node (along with some book-keeping
data) into a \SI{256}{\byte} flash page. This gives us a very shallow tree,
with a fanout of 81, which reduces the number of flash page accesses
considerably. For an example of the resulting node sizes (area of ground
covered), see \cref{tab:levels}. The last three columns show the number of
leaf nodes at the various levels in the experimental build described in
\cref{sec:exp}.

\begin{table}
\centering
\caption{Tree levels. Sizes are approximate}\label{tab:levels}
\def\m{\,\si{\meter}}
\begin{tabular}{
    c
    S[
        table-format=1.1e1,
        table-space-text-post=\m,
        table-align-text-post=false
    ]
    S[table-format=3]
    S[table-format=5]
    S[table-format=5]
}
\toprule
{\bf Level} & {\bf Size} & {\bf Zone} & {\bf VG} & {\bf Both}\\
\midrule
Top & 2.0e6\m & 0 & 0 & 0\\
1 & 2.1e5\m & 15 & 0 & 4\\
2 & 2.4e4\m & 625 & 168 & 471\\
3 & 2.5e3\m & 81 & 12353 & 39721\\
4 & 3.0e2\m & 0 & 157 & 486\\
5 & 3.0e1\m & 0 & 0 & 0\\
\bottomrule
\end{tabular}
\end{table}

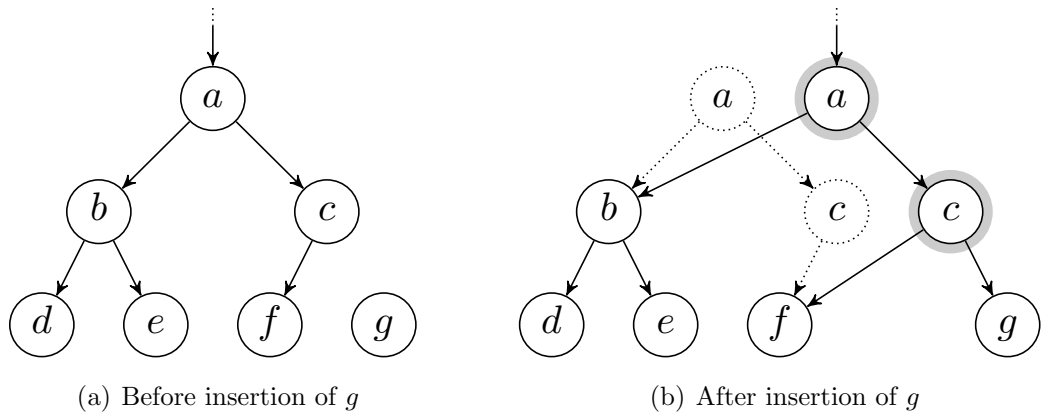
\begin{figure}
\def\Scale{1.5}
\centering
\subfigure[Before insertion of $g$]{\scalebox{\Scale}{\begin{tikzpicture}
    \fontsize{10}{10pt}\selectfont
    \begin{scope}[graph,directed]
    \draw
        (0,0) node (a) {\strut $a$}
        (-1,-1) node(b) {\strut $b$}
        (1,-1) node (c) {\strut $c$}
        (-1.5,-2) node (d) {\strut $d$}
        (-.5,-2) node (e) {\strut $e$}
        (.5,-2) node (f) {\strut $f$}
        (1.5,-2) node (g) {\strut $g$}

        (a) edge (b)
        (a) edge (c)
        (b) edge (d)
        (b) edge (e)
        (c) edge (f)
    ;
    \end{scope}
    \draw
        (a) ++(0,.8cm) coordinate (r)
        ($(r)!.3!(a.north)$) coordinate (r2)
        ;
    \draw
        (r) edge[densely dotted] (r2)
        (r2) edge[directed] (a)
        ;
\end{tikzpicture}}
\label{fig:insert:a}
}
\qquad
\subfigure[After insertion of $g$]{\scalebox{\Scale}{\begin{tikzpicture}
    \fontsize{10}{10pt}\selectfont
    \begin{scope}[graph,directed]
    \draw
        (0,0) node[densely dotted] (a) {\strut $a$}
        (-1,-1) node (b) {\strut $b$}
        (1,-1) node[densely dotted] (c) {\strut $c$}
        (-1.5,-2) node (d) {\strut $d$}
        (-.5,-2) node (e) {\strut $e$}
        (.5,-2) node (f) {\strut $f$}
        (1.5,-2) +(1,0) node (g) {\strut $g$}

        (a) +(1,0) node[hilite] (aa) {\strut $a$}
        (c) +(1,0) node[hilite] (cc) {\strut $c$}

        (a) edge[densely dotted] (b)
        (aa) edge (b)
        (a) edge[densely dotted] (c)
        (aa) edge (cc)
        (b) edge (d)
        (b) edge (e)
        (c) edge[densely dotted] (f)
        (cc) edge (f)
        (cc) edge (g)
    ;
    \end{scope}
    \draw
        (aa) ++(0,.8cm) coordinate (r)
        ($(r)!.3!(aa.north)$) coordinate (r2)
        ;
    \draw
        (r) edge[densely dotted] (r2)
        (r2) edge[directed] (aa)
        ;
\end{tikzpicture}}
\label{fig:insert:b}
}
\caption{Node $g$ is inserted by creating new versions of node $a$ and $c$
(highlighted), leaving the old ones (dotted) in place}\label{fig:insert}
\end{figure}

Immutable data structures have been used in purely functional programming for
decades~\citep[see, e.g.,][]{Okasaki:1999}, and they have recently become more
well known to the mainstream programming community through the data model used
in, for example, the Git version control system~\citep[see,
e.g.,][p.~3]{Loeliger:2012}. The main idea is that instead of modifying a data
structure in place, any nodes that would be affected by the modification are
duplicated. For a tree structure, this generally means the ancestor nodes of
the one that is, say, added. Consider the example in \cref{fig:insert}. In
\cref{fig:insert:a}, we see a tree consisting of nodes $a$ through $f$, and we
are about to insert $g$. Rather than adding a child to $c$, which is not
permitted, we duplicate the path up to the root, with the duplicated nodes
getting the appropriate child-pointers, as shown in \cref{fig:insert:b}, where
the duplicated nodes are highlighted. As can be seen, the old nodes (dotted)
are still there, and if we treat the old $a$ as the root, we still have access
to the entire previous version of the tree.

\section{Experimental Evaluation}\label{sec:exp}

\begin{figure}
\centering
\subfigure[VGs]{
\includegraphics[height=.5\textwidth]{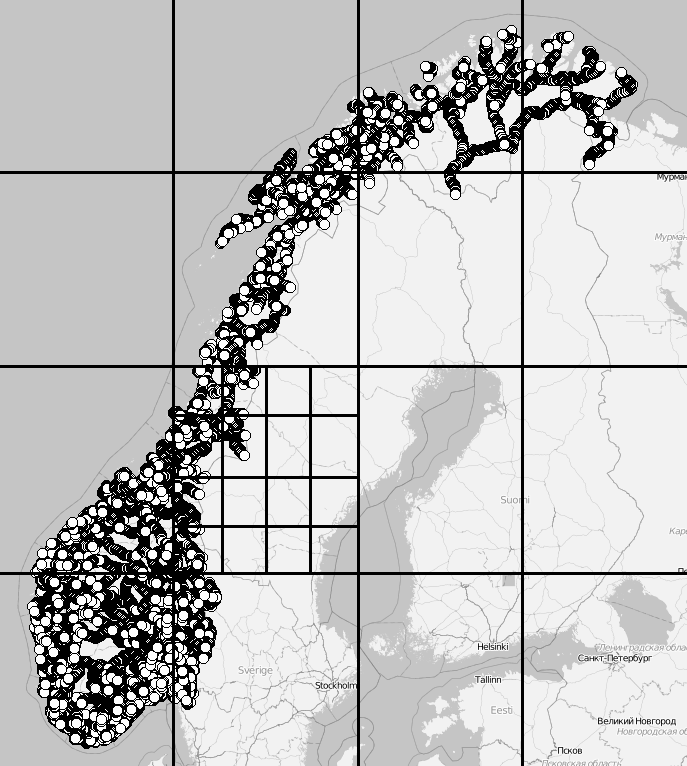}
\label{fig:gantriesno}
}
\hfill
\subfigure[Zones]{
\includegraphics[height=.5\textwidth, trim=0 0 22 0, clip]{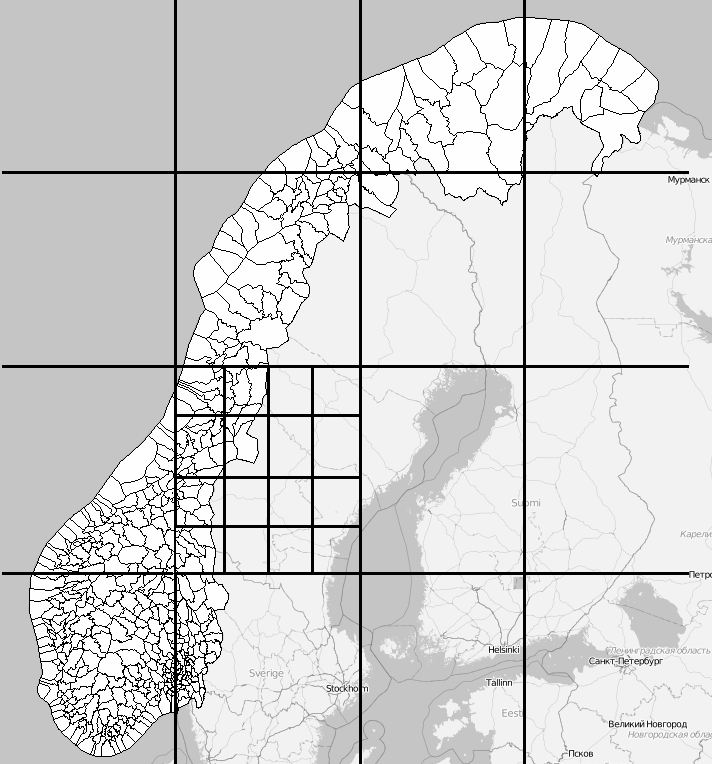}
\label{fig:zonesno}
}
\caption{Virtual gantries and zones of Norway, with an illustrative quadtree
grid}
\end{figure}

\begin{figure}
\centering
\subfigure[VGs]{
\includegraphics[height=.25\textwidth]{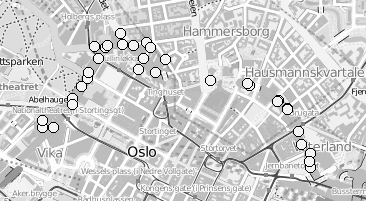}
\label{fig:gantriesosl}
}
\hfill
\subfigure[Zones]{
\includegraphics[height=.25\textwidth]{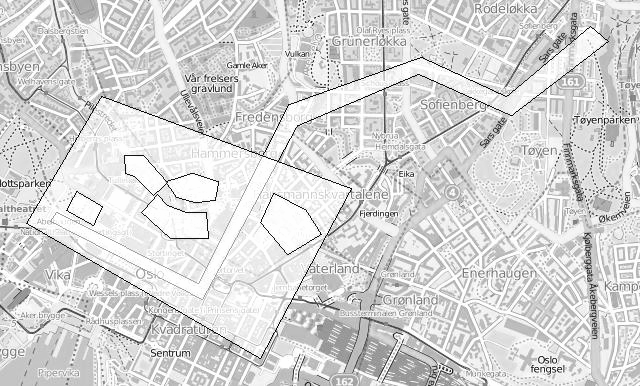}
\label{fig:zonesosl}
}
\caption{Virtual gantries and overlapping zones in downtown Oslo}
\end{figure}

Our experiments were performed with a data set of approximately \num{30000}
virtual gantries (see \cref{fig:gantriesno}), generated from publicly
available maps~\citep{Kartverket:2012}. The maps describe the main roads of
Norway with limited accuracy. Additionally, more detailed and accurate virtual
gantries were created manually for some locations in Oslo and Trondheim.
\Cref{fig:gantriesosl} shows the relevant virtual gantries in downtown Oslo,
used in our test drive. There are about 35 virtual gantries on this route, and
many of these are very close together. In general, there is one virtual gantry
before every intersection.

Each local administrative unit (\emph{kommune}) is present in the maps
used~\citep{Kartverket:2012}, with fairly accurate and detailed boundaries.
There are \num{446} such zones in total (see \cref{fig:zonesno}). In addition,
more detailed and accurate zones were created manually for some locations in
the cities of Oslo and Trondheim. Some of these are quite small, very close
together, and partially overlapping (see \cref{fig:zonesosl}).

\subsection{Data Structure Build}

\begin{figure}
\centering
\begin{tikzpicture}
    \begin{axis}[
            width=10cm,
            height=7cm,
            ybar,
            ylabel={Flash pages},
            enlarge x limits=0.22,
            symbolic x coords={Zones, VGs, Both},
            xtick=data,
        ]
        \addplot [
            fill=white
        ] coordinates {
            (Zones, 2164)
            (VGs, 42217)
            (Both, 71675)
        };
        \label{plot:flash}
        \addplot [
            pattern=crosshatch dots
        ] coordinates {
            (Zones, 1716)
            (VGs, 13180)
            (Both, 42190)
        };
        \label{plot:idxflash}
        \addplot [
            pattern=crosshatch
        ]
        coordinates {
            (Zones, 2021)
            (VGs, 41147)
            (Both, 45131)
        };
        \label{plot:flashnodups}
        \addplot [
            fill=black
        ]
        coordinates {
            (Zones, 1573)
            (VGs, 12110)
            (Both, 15646)
        };
        \label{plot:idxflashnodups}
\end{axis}
\end{tikzpicture}
\caption{The plot shows total flash pages used (\plotref{plot:flash})
    and flash pages used for the index
    (\plotref{plot:idxflash}), as well as the same with duplicates removed
    (\plotref{plot:flashnodups} and \plotref{plot:idxflashnodups},
    respectively) for a data base consisting of zones, virtual gantries, or
    both (c.f., \cref{tab:perf})}\label{fig:memuse}
\end{figure}
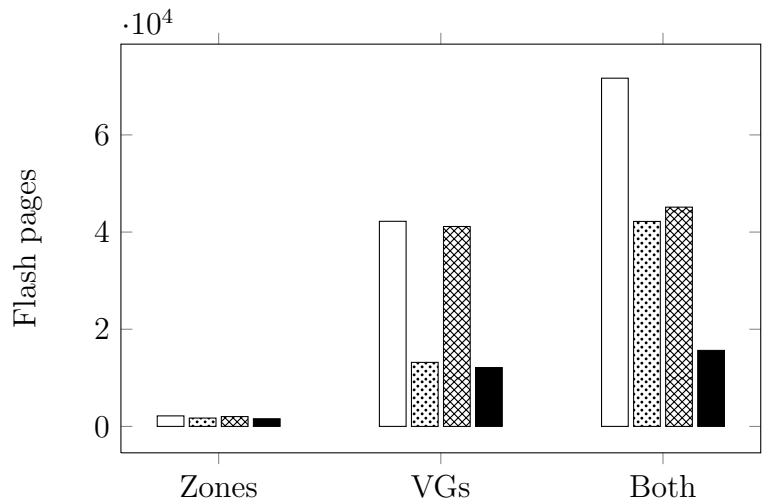

These test data were inserted into the quadtree as described in
\cref{sec:solution}. \Cref{tab:perf} summarizes the important statistics of
the resulting structure. The numbers for memory usage are also shown in
\cref{fig:memuse}, for easier comparison.

\rowcolors{1}{white}{gray!10}
\ra{1.15} \begin{table}
    \caption{Tree performance numbers}\label{tab:perf}

\def\MB{\,\si{\mebi\byte}}
\begin{tabularx}{\textwidth}{
                >{\it}c
        X
        S[
            table-format=4.2,
                                ]
        S[
            table-format=5.2,
                                ]
        S[
            table-format=5.2,
                                ]
            }
    \toprule
    \rowcolor{white}
    & \bf Description & {\bf Zones} & {\bf VGs} & {\bf Both}\\
    \specialrule{\lightrulewidth}{2pt}{0pt}     a & Number of objects in database & 448 & 29037 & 29485 \\
b & Flash pages for index and data & 2164 & 42217 & 71675 \\
c & Size (\si{\mebi\byte}) & 0.53 & 10.31 & 17.50 \\
d & Flash pages for index & 1716 & 13180 & 42190 \\
e & Objects referenced by leafs & 2481 & 43263 & 95272 \\
f & Leaf nodes per object & 5.5 & 1.5 & 115 \\
g & Leaf entries not used, empty & 240 & 27403 & 1319 \\
h & Leaf entries set & 733 & 13179 & 41207 \\
i & Max index tree depth & 3 & 4 & 4 \\
j & Zone inside entries & 57 &  & 30379 \\
k & Zone edge entries & 2406 &  & 21333 \\
l & Distinct leaf pages & 578 & 11608 & 14138 \\
m & Total leaf pages & 721 & 12678 & 40682 \\
n & Duplicate leaf pages & 143 & 1070 & 26544 \\
o & Flash pages, dups removed & 2021 & 41147 & 45131 \\
p & Size, dups removed (\si{\mebi\byte}) & 0.49 & 10.05 & 11.02 \\
q & Index pages, dups removed & 1573 & 12110 & 15646 \\
r & Size of index, dups removed (\si{\mebi\byte}) & 0.38 & 2.96 & 3.82 \\

\bottomrule
\end{tabularx}
\end{table}

Judging from these numbers (row \textit{r}), a database containing only zones
would be quite small (about \SI{0.38}{\mebi\byte}). Each zone is referenced by
\num{5.5} leaf nodes on average (row \textit{f}). Also note that only \num{57}
leaf nodes (squares) are entirely contained in a zone (row \textit{j}). This
implies that the geometric inside/outside calculations will need to be
computed in most cases.

This can be contrasted with the combined database of gantries and zones. The
index is larger (\SI{3.82}{\mebi\byte}, row \textit{r}), but the performance
of polygon assessments is much better. Each polygon is referenced from
\num{115} leaf nodes (row \textit{f}) and there are more \emph{inside} entries
than \emph{edge} entries (\num{30379} vs \num{21333}). This indicates that the
geometric computations will be needed much less frequently.

Each of the three scenarios creates a number of duplicate leaf pages. Many
leaf pages will contain the same zone edge/inside information. In our
implementation of the algorithm, this issue is not addressed or optimized. It
is, however, quite easy to introduce a reference-counting scheme or the like
to eliminate duplicates, in this scenario saving \SI{6}{\mebi\byte} (as shown
in rows \textit{o} through \textit{r}).

The zone database contains very few empty leaf entries (row \textit{g}),
because the union of the regions covers the entire country, with empty regions
found in the sea or in neighboring countries.

\subsection{Flash Access in a Real-World Scenario}

\begin{figure}
\centering
\includegraphics[width=\textwidth, trim=10 0 10 0, clip]{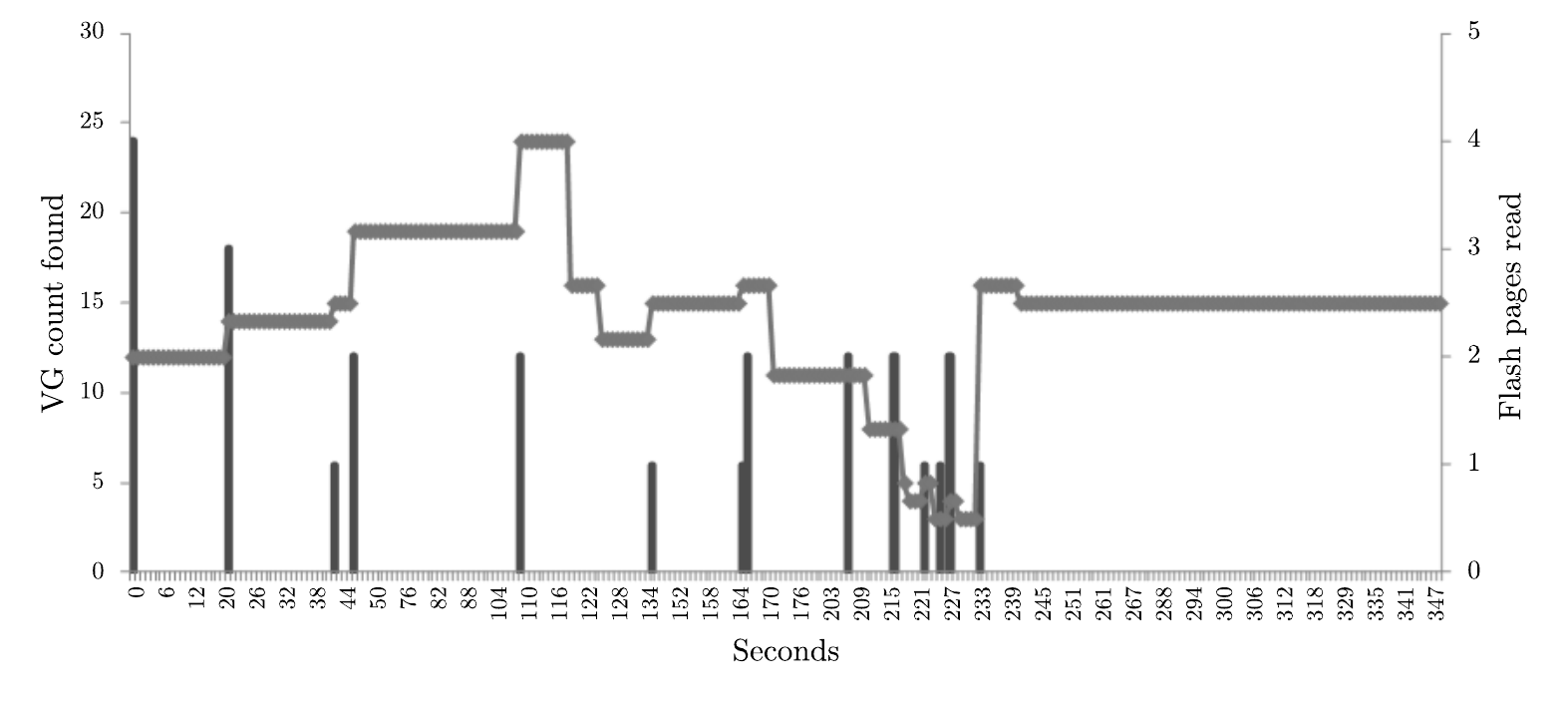}
\caption{Flash access in an actual drive (Oslo Ring-1 Eastbound): flash pages
    read (vertical bars) and virtual gantries found (horizontal
    lines)}\label{fig:access}
\end{figure}

The index was also tested in a \SI{2}{\kilo\meter} drive, eastbound on
Ibsenringen, in downtown Oslo. The relevant virtual gantries are shown in the
map in \cref{fig:gantriesosl}. The in-memory flash cache used was 15 pages
($\num{15}\cdot\SI{256}{\byte}$), and the cache was invalidated before test
start. \Cref{fig:access} shows the result, in terms of flash accesses and the
number of gantries found.

\section{Discussion} 

To view our results in the context of the initial problem, we revisit our
requirement list from \cref{sec:problem}. We can break our solution into three
main features:
\begin{enumerate*}[label=(\textit{\roman*}), itemjoin={{; }},
        itemjoin*={{; and }}]
    \item The use of quadtrees for indexing
    \item purely functional updates and immutability
    \item high fanout, with a 9-by-9 grid.
\end{enumerate*}
\Cref{tab:satisfy} summarizes how these features, taken together, satisfy all
our requirements. Each feature is either partly or fully relevant for any
requirement it helps satisfy (indicated by \P or \Y, respectively).

Our starting-point is the need for spatial (two-dimensional) indexing
(\cref{req:oper}), and a desire for simplicity in our solution. The slowness
of our hardware made a straightforward linear scan impossible, even with a
data set of limited size. This led us to the use of quadtrees, whose primary
function, seen in isolation, is satisfying \cref{req:cpu}, CPU efficiency. It
also supports \cref{req:foot} (low memory footprint) by giving us a platform
for reducing duplication. Lastly, it supports efficiency in terms of flash
page accesses (\cref{req:flash}), which is primarily handled by high fanout,
using the nine-by-nine grid.

The purely functional updates, and the immutable nature of our structure,
satisfies a slew of requirements by itself. Just as in modern version control
systems such as Git~\citep{Loeliger:2012}, immutable tree structures where
subtrees are shared between versions gives us a highly space-efficient way of
distributing and storing multiple, incremental versions of the database
(\cref{req:vers,req:incr,req:foot}). This also gives us the ability to keep
using the database during an update, and to roll back the update if an error
occurs, without any impact on the database use, as the original database is
not modified (\cref{req:itgr}). Finally, because modifications will always use
new flash pages, we avoid excessive modifiations of, say, the root node, and
can schedule the list of free nodes to attain a high degree of wear leveling
(\cref{req:wear}).

In our tests, as discussed in the previous section, we found that the solution
satisfied our requirements not only conceptually, but also in actual
operation. It can contain real-world data within real-world memory constraints
(\cref{tab:perf}), and can serve up results in real time during actual
operation, with relatively low flash access rates (\cref{fig:access}).

\rowcolors{1}{}{}
\begin{table}
    \centering
    \caption{How components of the solution satisfy various requirements}
                \label{tab:satisfy}
    \begin{tabularx}{\textwidth}{@{}Xcccccccc@{}}
        \toprule
        \bf Feature
          & \bf A & \bf B & \bf C & \bf D & \bf E & \bf F & \bf G & \bf H \\
                \midrule
        The use of quadtrees for indexing
          & \N & \N & \P & \N & \Y & \P & \N & \Y \\
        Purely functional updates and immutability
          & \Y & \Y & \Y & \Y & \N & \N & \Y & \N \\
        High fanout, with 9-by-9 grid
          & \N & \N & \N & \N & \N & \Y & \N & \N \\
        \bottomrule
    \end{tabularx}
\end{table}

\section{Conclusion \& Future Work}

We have described the problem of real-time spatial indexing in an in-vehicle
satellite navigation (\textsc{gnss}) unit for the purposes of open-road
tolling. From this problem we have elaborated a set of performance and
functionality requirements. These requirements include issues that are not
commonly found in indexing for ordinary computers, such as the need for
wear-leveling over memory locations. By modifying the widely used quadtree
data structure to use a higher fanout, and by making it immutable, using
purely functional updates, we were able to satisfy our entire list of
requirements. We also tested the solution empirically, on real-world data and
in a real-world context of a vehicle run, and found that it performed
satisfactorily.

Although our object of focus has been a rather limited family of hardware
architectures, the simple, basic ideas of our index chould be useful also
for other devices and applications where real-time spatial indexing is required
under somewhat similar flash memory conditions. Possible extensions of our
work could be to test the method under different conditions, perhaps by
developing a simulator for in-vehicle units with different architectures and
parameters. This could be useful for choosing among different hardware
solutions, as well as for tuning the index structure and database.

\paragraph{Disclaimer \& Acknowledgements}
Magnus Lie Hetland introduced the main design idea of using immutable,
high-fanout quadtrees for the database structure. He wrote the majority of the
text of the current paper, based in large part on the technical report of
\citet{Lykkja:2012}.
Ola Martin Lykkja implemented and benchmarked the database structure and
documented the experiments~\citep{Lykkja:2012}.
Neither author declares any conflicts of interest. Both authors have revised
the paper and approved the final version. The authors would like to thank Hans
Christian Bolstad for fruitful discussions on the topic of the paper. This
work has in part been financed by the Norwegian Research Council (BIA project
no.~210545, ``SAVE'').

\bibliography{paper}
\end{document}